\documentclass[conference]{IEEEtran}

\usepackage{eurosym}
\usepackage{amssymb}
\usepackage{amsmath}
\usepackage{graphicx}
\begin{document}
\title{SQL Access Patterns for Optimistic Concurrency Control}

\author{\IEEEauthorblockN{Fritz Laux}
\IEEEauthorblockA{Fakult\"at Informatik\\
Reutlingen University\\
D-72762 Reutlingen, Germany\\
fritz.laux@reutlingen-university.de}
\and
\IEEEauthorblockN{Martti Laiho}
\IEEEauthorblockA{Dept. of Business Information Technology\\
Haaga-Helia University of Applied Sciences\\
FIN-00520 Helsinki, Finland\\
martti.laiho@haaga-helia.fi}
}

\maketitle
\footnote{This paper is the result of collaborative work undertaken along the lines of the DBTechNet Consortium. The authors participate in DBTech EXT, a project partially funded by the EU LLP Transversal Programme (Project Number: 143371-LLP-1-2008-1-FI-KA3-KA3MP)}
\begin{abstract}
Transaction processing is of growing importance for mobile and web applications. Booking tickets, flight reservation, e-Banking, e-Payment, and booking holiday arrangements are just a few examples. Due to temporarily disconnected situations the synchronisation and consistent transaction processing are key issues. To avoid difficulties with blocked transactions or communication loss several authors and technology providers have recommended to use Optimistic Concurrency Control (OCC) to solve the problem. 
However most vendors of Relational Database Management Systems (DBMS) implemented only locking schemes for concurrency control which prohibit the immediate use of OCC.
We propose Row Version Verifying (RVV) discipline to avoid lost updates and achieve a kind of OCC for those DBMS not providing an adequate non-blocking concurrency control. Moreover, the different mechanisms are categorized as access pattern in order to provide programmers with a general guideline for SQL databases. 
The proposed SQL access patterns are relevant for all transactional applications with unreliable communication and low conflicting situations.
We demonstrate the proposed solution using mainstream database systems like Oracle, DB2, and SQLServer.
\end{abstract}
\IEEEpeerreviewmaketitle

%------------------------------------------------------------------------- 
\section{Introduction}
Mobile applications enable users to execute business transactions while being on the move. 
It is essential that temporary disconnected situations do not compromise transaction properties or block database resources on the server. To prevent blocked resources researchers have intensively studied OCC \cite{r1, r2}, but no commercial database product has implemented this mechanism, yet.
With the popularity of multitier software architectures technology vendors like those for J2EE platforms, object relational mappers or Service Oriented Architecture (SOA) have proposed to use OCC to solve the problem.

But shifting the burden to the middleware is a tricky task. The designer and implementer of a transactional application have to leave the DBMS unaware of the user transaction to avoid the automatic locking of data for an unpredictable time. On the other hand, concurrent transactions of different applications may interfere without the possibility for any help by the DBMS. Therefore the applications and the DBMS need to co-operate somehow to ensure that at least the lost update problem will be avoided.

A typical fault in multi-user file-based systems without proper concurrency control is the lost update problem i.e. a record $x$ updated by some process A will be overwritten by some other concurrent process B like in the following problematic canonical schedule \cite[pp. 62-63]{r4}: $r_A(x), r_B(x), w_A(x), w_B(x)$, where $r_T(x)$ and $w_T(x)$  denote read and write operations of transaction T on data item $x$  .

A properly used DBMS would not allow such a situation to happen because it would lock $x$ for transaction A and prevent B from accessing $x$ before A commits or aborts. But, if the database does not receive a termination request, e.g. because of a communication failure, the record $x$ remains blocked. 

We do not want to risk blocked data, therefore a kind of OCC should be applied. Even if the DBMS does not support OCC directly we will show that it could help the application to detect concurrency conflicts.    
For relational databases we will show how this can be achieved using a row version column for every table and specific access patterns. %essentially propose to add a row version column for every table in order to easily detect any data changes.
\subsection{Structure of the Paper}
After a motivation for our approach and the related work we present in Section \ref{sec:LostUpd} the lost update problem by example.  Section \ref{sec:Taxonomy} describes three SQL patterns that solve the problem and in Section \ref{sec:ServerStamping} we provide an implementation for a server side row version column to support OCC for mainstream SQL databases. In Section \ref{Conclusion} we conclude our findings.

\subsection{Motivation}
Kung and Robinson \cite{r1} distinguishes three phases of a transaction when OCC is used:
\begin{itemize}
\item read phase
\item validation phase
\item write phase
\end{itemize}

The first phase includes user input and thinking time. It may last for an unpredictable time span. The following phases are without any user interaction. Validation and write phases are therefore very short in the range of milliseconds. The last two phases are critical in the sense that exclusive access is required. Failing to do so could result in inconsistent data, e.g lost update. A Relational Database Management System (RDBMS) could help to support each phase by choosing the proper transaction isolation level.  The read phase should read only valid data (READ COMMITTED) and transaction mode can be set to READ ONLY. Switching to a strong enough isolation level (REPEATABLE READ, SNAPSHOT, or SERIALIZABLE) during validation and write phases will yield the corresponding transaction properties against competing transactions (see Fig. \ref{fig:OCCpatterns}).

The transaction isolation level may only be altered before or as the first statement of a transaction. This implies that our user transaction has to be split up into two database/SQL transactions. Each SQL transaction should be set to the isolation levels as recommended before. During the validation phase the application has to re-read the data and check for any changes by concurrent transactions. If any changes are detected, then the transaction has to abort else it may proceed with the write phase. 

Applying OCC to the previous example the result would be the following history: $r_A(x), r_B(x), val_A, w_A(x), val_B, a_B$, where $val_A$ denotes the validation of transaction $A$ and $a_B$ denotes the abort command for transaction $B$. 

Or, if transaction A decides to abort then B will be successful with $r_A(x), r_B(x), a_A, val_B, w_B(x)$.  In either case only one of the competing transactions can be successful.

This example also shows that the non blocking concurrency control comes for the prize of transaction aborts. 

Instead of optimistic concurrency control theories presented in database textbooks, we are interested in ways how to implement such a mechanism using mainstream DBMS systems and what application developers need to understand about reliable access of databases. 

We therefore extend the read-write model as used by Herbrand's semantic  (see \cite{r4,r14,r15}) to fit with the OCC mechanism. If the validation is not passed successfully then the write phase will be skipped, leading to an aborted transaction. Instead of having validation $val_T(x)$ and write $w_T(x)$ we introduce a conditional write operation $w(x, k)$. This write operation on the data item $x$ is only executed if the condition $k$ evaluates to true. Checking $k$ may require to read the actual database state. Reading, checking $k$, and writing $x$ do not allow any parallel operations as explained above. 

In this paper we present access patterns that implement this $w(x,k)$ operation. A simple example of implementation would use solely the SQL update command: 
\begin{center}
\textsc{UPDATE} $table$  \textsc{SET} $X$ = $val$  \textsc{WHERE} $k$ \textsc{AND} $id(x)$
\end{center}
where $id(x)$ evaluates to true only for the row of data item $x$.

\subsection{Related Work}
Concurrency control is a cornerstone of transaction processing, it has been extensively studied for decades. Namely Gray and Reuter \cite{r3} studied locking schemes, whereas Kung and Robinson \cite{r1} developed optimistic methods for concurrency control. Unland \cite{r2} presents OCC algorithms without critical section. Using these algorithms would allow relaxed isolation levels but involve checking the read set against all concurrent transactions. Because the application is not aware of concurrent transactions its use can be ruled out in our case.

A higher concurrency for query intensive transactions provide Multiversion Concurrency Control (MVCC) as described by Stearns and Rosenkrantz \cite{r18} and Bernstein and Goldman \cite{r19}. If we check the MVCC method for its usability for web or mobile transaction processing it is even worse than locking in terms of resource consumption. While locking needs to record only the id of the item locked, the MVCC needs to store a version each time an item is updated that was read by an active transaction prior to the update. In case of disconnected situations this may lead to large number of versions for a single data item.

With the dissemination of middleware OCC has been recommended by IT-vendors (\cite{r20,r21,r22}) for transactional e-business and m-commerce applications but little concern have been spend on how this can be achieved using commercial SQL databases \cite{r7}. He\ss \cite{r20} simply uses Hibernate's optimistic-lock="version" option but does not mention the risk of legacy applications not under control of Hibernate which could still lead to lost updates. Nock \cite{r6} uses a timestamp column with Java timestamp resolution ignoring the fact that contemporary database products can produce more than a hundred times the same timestamp \cite{r7}.  
Akbar-Husain \cite{r21} believes that demarking the method that checks the version with the required transaction attribute will be sufficient to avoid lost updates. He fails to tell that only a strong enough isolation level will achieve the desired results. 

\section{Lost Update Problem in the Application Context}
\label{sec:LostUpd} 
Let us consider first the following problematic scenario of SQL transactions of two concurrent processes A and B updating the balance of the same account in Table \ref{tab:lostUpd}. 

\begin{table}[]
\caption{A lost update scenario using SELECT-UPDATE in transaction A}
\label{tab:lostUpd}
\bigskip
\begin{tabular}{r|l|r|l}
step & process A & balance & process B\\
\hline
1 & \textsc{SET TRANSACTION} &  & \\
   & \textsc{ISOLATION LEVEL}   &   & \\
   & \textsc{READ COMMITTED}  &   & \\
\hline
2 &   &  1000\EUR & \\
\hline
3 & \textsc{SELECT balance} &  & \\
   & \textsc{INTO :balance}   &   & \\
   & \textsc{FROM Accounts}  &   & \\
   & \textsc{WHERE acctId = :id;}  &   & \\
\hline
4 &   &   & \\
\hline
5 & \textsc{newBalance =} &  & \textsc{UPDATE Accounts}\\
   & \textsc{balance - 100} &   &  \textsc{SET balance =}  \\
   & &   &  \textsc{balance - 200}   \\
   & &   &  \textsc{WHERE acctId = :id;}  \\
\hline
6 &   &  800\EUR & \textsc{COMMIT;}\\
\hline
7 & \textsc{UPDATE Accounts} &  & \\
   & \textsc{SET balance =}   &   & \\
   & \textsc{:newBalance}  &   & \\
   & \textsc{WHERE acctId = :id;}  &   & \\
\hline
8 & \textsc{COMMIT;}  &  900\EUR & \\
\hline
\end{tabular}\\
\end{table}

% \begin{figure}[!t]
%\centering
%\includegraphics[width=0.35\textwidth]{lostUpdate}
%\caption{A lost update scenario using SELECT-UPDATE in transaction A}
%\label{fig_lostUpd}
%\end{figure}

The withdrawal of 200  made by the transaction of B will be overwritten by A, in other words the update made by B in step 5 will be lost in step 7 when the transaction of A overwrites the updated value by value 900  which is based on stale data i.e. outdated value of the balance from step 3.  If the transactions of A and B serialized properly, the correct balance value after these transactions would be 700, but there is nothing that the DBMS could do to protect the update of step 5 since the guilty party to this lost update problem is the programmer of process A, who has ordered a wrong isolation level from the DBMS. READ COMMITTED, which for performance reasons is the default transaction isolation level used by most RDBMS systems, does not protect any data read by transaction of getting outdated right after reading the value.  Locking Scheme Concurrency Control (LSCC) prevents conflicting access to data. Conflicts are defined in terms of isolation levels. The proper isolation level on LSCC systems to prevent a lost update should be REPEATABLE READ or SERIALIZABLE, which would protect the values read in the transaction from getting outdated during the transaction by holding shared locks on these rows up to the end of the transaction.  The isolation service of the DBMS does guarantee that the transaction will either get the ordered isolation or, in case of serialization conflict, the transaction will be rejected by the DBMS.  The means used for this service and the transactional outcome for the very same application code can be different when using different DBMS systems, and even in using different table structures. A LSCC may as well delay to grant a lock request until the possible conflict disappears. Usually a transaction rejected due to a serialization conflict should be retried by the application, but we will discuss this later.

The erroneous scenario above would also be the same if process A commits its transaction of steps 1 and 3 (let us call it transaction A1) in step 4, and continues (for example after some user interaction) with another transaction A2 of phases 7-8.  In this case, no isolation level can help, but transaction A2 will make a blind write (based on stale data, insensitive of the current value) over the balance value updated by transaction B.

\section{SQL Access Patterns for Avoiding Lost Updates}
\label{sec:Taxonomy}
\begin{figure}[]
\centering
\includegraphics[width=0.4\textwidth]{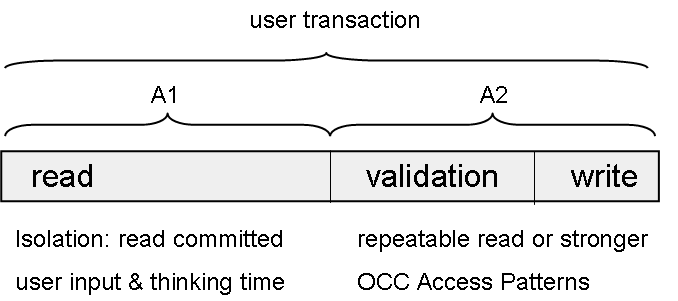}
\caption{Context of the OCC Access Patterns}
\label{fig:OCCpatterns}
\end{figure}

The blind write of the update transaction A2 of phases 7-8 (resulting in the lost update of transaction B) could have been avoided by any of the following practice. The access patterns apply to the validation and write phase (process A2) as shown in Figure \ref{fig:OCCpatterns}.
 We present the patterns in the canonical form given by Coplien \cite{r23} that is shorter and more essential than the one used by Gamma et al \cite{r5}:
\subsection{Access Pattern: Sensitive UPDATE}
\begin{description}
\item [\emph{Problem}:] \quad How to prevent a lost update in case of concurrent updates without using explicit locks.
\item [\emph{Context}:] \quad Concurrent transaction processing in distributed systems has to deal with temporary disconnected situations and nevertheless ensure correct results.
\item [\emph{Forces}:] \hspace{1mm} 
\end{description}
\begin{itemize}
\item Using locks to prevent other transactions from changing the value can block data items for unpredictable time in case of communication failure or in case of long user thinking time.
\item Multiversion concurrency control (MVCC) or OCC do not block data access, but lead to abort conflicting transactions except for the first one that updates the data.
\item OCC is not supported by commercial SQL databases, hence we cannot directly use DBMS support.
\end{itemize}
\begin{description}
\item [\emph{Solution}:] \quad There is no risk of lost update if A2 in step 7 uses the form of the update which is sensitive to the current value, like B uses in step 5 as follows:
\begin{quote}
UPDATE Accounts\\
SET balance = balance - 100\\
WHERE acctId = :id;
\end{quote}
\item [\emph{Consequences}:] \hspace{11mm} It should be noted that the update of the balance is based on a value that is not seen by the application and therefore the user will not be aware of the changed balance. So, this access pattern does not provide repeatable read isolation.  If the user needs to know about the changed situation the access pattern "Re-SELECT .. UPDATE" could be used (see below).
\end{description}
\subsection{Access Pattern: Conditional UPDATE}
\begin{description}
\item [\emph{Problem}:] \quad How to prevent a lost update and provide repeatable-read for a user transaction in case of concurrent updates without using locking. 
\item [\emph{Context}:] \quad The "Sensitive UPDATE" pattern in concurrent read situations may result in non-repeatable phenomenon.
\item [\emph{Forces}:] \hspace{1mm}  Same as for "Sensitive UPDATE" plus:
\end{description}
\begin{itemize}
\item The data value read and displayed to the user may not be the same on which the update is based (non-repeatable read phenomenon).
\end{itemize}
\begin{description}
\item [\emph{Solution}:] \quad 
After transaction A1 first has read the original row version data in step 3, transaction A2 verifies in step 7, using an additional comparison expression in the WHERE clause of the UPDATE command, that the current row version in the database is still the same as it was when the process previously accessed the account row, for example, 
\begin{quote}
UPDATE Accounts\\
SET balance = :newBalance \\
WHERE acctId = :id AND \\
 (rowVersion = :old\_rowVersion);
\end{quote}
The comparison expression can be a single comparison predicate like in the example above where rowVersion is a column (or a pseudo-column provided by the DBMS) reflecting any changes made in the contents of the row and :old\_rowVersion is a host variable containing the value of the column when the process previously read the contents of the row. In the case that more than one column is involved in the comparison, the expression can be built of version comparisons of all columns used and based on the 3-value logic of SQL.
\item [\emph{Consequences}:] \hspace{11mm} 
Since this access pattern does not explicitly read data, there is no need to set isolation level.  The result of the concurrency control services is the same for locking scheme concurrency control (LSCC) and multiversion concurrency control (MVCC) based DBMS.  The result of the update depends on the row version verifying predicate, and the application code needs to evaluate the return code to find out the number of updated rows to verify the result.
\end{description}
\subsection{Access Pattern: Re-SELECT .. UPDATE} 
\begin{description}
\item [\emph{Problem}:] \quad How to provide repeatable-read for a user transaction in case of concurrent updates without using locks. Signal the user if conflicting transactions have changed the read set. 
\item [\emph{Context}:] \quad "Conditional UPDATE" pattern does not allow to inform the user of the changed read set before aborting the transaction.  
\item [\emph{Forces}:] \hspace{1mm}  Same as for "conditional UPDATE" plus:
\end{description}
\begin{itemize}
\item In the time span between the re-SELECT and the UPDATE statement the data read may be updated again by concurrent transactions. In the worst case, this can lead to an infinite loop.
\item Executing the pattern in repeatable read isolation may force the transaction to abort if no locking is used.
\end{itemize}
\begin{description}
\item [\emph{Solution}:] \quad 
This is a variant of the "conditional UPDATE" pattern in which transaction A2 first reads the current row version data from the database into some host variable current\_rowVersion which allows the application to inform the user of the changed situation:
\begin{quote}
SELECT rowVersion \\
INTO :current\_rowVersion\\
FROM Accounts\\
WHERE acctId = :id; \\
// ... inform the user if desired
\end{quote}
and then apply the conditional update:
\begin{quote}
if (current\_rowVersion = old\_rowVersion) then\\
\hspace*{3 mm} UPDATE Accounts\\
\hspace*{3 mm} SET balance = :newBalance \\
\hspace*{3 mm} WHERE acctId = :id ;
\end{quote}
To avoid repeatedly re-SELECT, it is necessary to make sure that no other transaction can change the row between the SELECT and the UPDATE.  For this purpose, we need to apply a strong enough isolation level (REPEATABLE READ, SNAPSHOT, or SERIALIZABLE) or explicit row-level locking, such as Oracle's FOR UPDATE clause in the SELECT command. 
\item [\emph{Consequences}:] \hspace{11mm} 
Since isolation level implementations of LSCC and MVCC based DBMS are different, the result of concurrency services can be different:  In LSCC based systems the first writer of the row or reader using REPEATABLE READ or SERIALIZABLE isolation level will usually win, whereas in MVCC based systems the first writer wins the concurrency competition.
\end{description}
\section{RVV Discipline and Server Side Stamping}
\label{sec:ServerStamping}
The last access pattern doesn't require any locking before transaction step 7 (start of A2). This update method is generally known as "Optimistic Locking" \cite{r6}, but we prefer to call it Row Version Verification (RVV) Discipline. There are multiple options for row version verification, including comparison of original contents of all or some relevant subset of columns of the row, a checksum of these, a technical SQL column, or some technical pseudo-column maintained by the DBMS.

A general solution for row version management is to include a technical row version column $rv$ and to use a row-level trigger to increase the value of column $rv$ on any row automatically every time the row is updated.  We call the use of trigger or use of technical pseudo-column as "server-side stamping" which no application can bypass, as opposite to client-side stamping using the SET clause within the UPDATE command - a discipline that all applications should follow in that case.
Row-level triggers are affordable, but have performance cost of some percents in execution time on Oracle and DB2, whereas SQL Server does not even support row-level triggers.

Timestamps are typically mentioned in database literature as a means of differentiating any updates of a row. However, our tests \cite{r7} prove that, for example, on a 32bit Windows workstation using a single processor Oracle 11g can generate up to 115 updates having the very same timestamp. Almost the same problem applies to DATETIME of SQL Server 2005 and TIMESTAMP of DB2 LUW 9, with exception of the new ROW CHANGE TIMESTAMP option in DB2 9.5 which generates unique timestamp values for every update of the same row having technical TIMESTAMP column.

The native TIMESTAMP data type of SQL Server is not a timestamp but a technical column which can be used to monitor the order of all row updates inside a database.  We prefer to use its synonym name ROWVERSION. This provides the most effective server-side stamping method in SQL Server; although, as a side-effect it generates an extra U-lock which will result in a deadlock in the example of Figure \ref{tab:lostUpd}.

In version 10 and later versions, Oracle provides a new pseudo-column ORA\_ROWSCN for rows in every table created with the ROWDEPENDENCIES option \cite{r8}.  This will show the transaction's System Change Number (SCN) of the last committed transaction which has updated the row.  This provides the most effective server-side stamping method for RVV in Oracle databases, although as a harmful side-effect, the row-locking turns its value to NULL.

In our "RVV Paper" \cite{r7}, we have presented an SQL view as solutions for mapping these technical row version column contents into BIGINT data type for Row Version Verification (RVV) at the client-side.

\section{Conclusion}
\label{Conclusion}
The concurrency control by DBMS treats SQL transactions without their application context in line with Herbrand semantics, and this is the typical scope of database textbooks in teaching transaction programming.  We see the need to expand this scope to the application level, to typical user transactions which are the context for SQL transactions.  Even if the widely accepted Design Patterns of GoF \cite{r5} do not even mention database transactions, we can identify and build practical Data Access Patterns to be used for teaching Data Access Technologies.  

Modern application architectures have introduced new practices and needs which have outdated some practices of earlier SQL programming like locking and holdable cursors. Commercial database management systems do not yet support OCC which is needed for mobile and web-applications. So, for example, we had to develop access patterns for optimistic locking services on the user level. We presented three of these patterns and showed how far current DBMS can support it.

\end{document}